\documentclass[preprint,aps]{revtex4-1}
\usepackage{graphicx,color}

\begin{document}

\title{Drop Dynamics of Viscoelastic Filament}
\author{Hrishikesh Pingulkar,$^1$ Jorge Peixinho,$^{1,2}$ and Olivier Crumeyrolle$^1$}
\affiliation{$^1$Laboratoire Ondes et Milieux Complexes, CNRS and Universit\'{e} Le Havre Normandie, 76600 Le Havre, France \\
$^2$Laboratoire PIMM, CNRS, Arts et M\'{e}tiers Institute of Technologie, Cnam, HESAM Universit\'{e}, 75013 Paris, France}
\date{\today}

\begin{abstract}
The stretching of viscoelastic polymer solutions close to break-up can create attached drops on a filament, whose properties and dynamics are little understood. 
The  stretching of capillary bridges and the consecutive filament, until its breakup, can be quantified using diameter-space-time diagrams, which demonstrate hierarchy, as well as, asymmetry of satellite drops around a big central drop. All drops experience migration, oscillation and merging. 
In addition, the position of the minimum diameter on the filament is determined, along with the number of drops, their positions, the diameters of drops and the filament breakup time.
The maximum number of drops on the filament can be predicted using the Deborah number.
The diagrams also quantify the large Hencky strains in the filaments before pinch-off.
The obtained minimum diameter is used to measure the extensional viscosity, which indicates the effect of polymer concentration and direction of filament thinning.
\end{abstract}

\maketitle

\section{Introduction}
Understanding the extensional flow properties of polymer solutions is of practical and physical importance for many commercial applications such as spraying, coating, inkjet printing, food processing, atomization, etc. 
Most of these processes undergo filament breakup of solutions containing dissolved polymers and the extensional viscosity of these solutions plays an important role in the thinning and the drop dynamics.
In contrast to Newtonian fluids, which have their extensional viscosity directly proportional to the shear viscosity, the extensional viscosity of viscoelastic fluids is more complex. 
Macromolecular solutions exhibit large extensional viscosity \cite{Sridhar1991,Dinic2015,McKinley2005,Bergeron2000,Keller1985,Sharma2015,Campo-Deano2010,McKinley2002,Clasen2006,Gaillard2019} because extensional flows are irrotational and presumably more efficient at disentangling or orienting flexible polymer molecules. 
It has been known for 50 years \cite{Middleman1965,Goldin1969} that capillary jets of viscoelastic polymer solutions exhibit the peculiar morphology called beads-on-a-string (BOAS). The instability and the initial sinusoidal growth has been reported \cite{Clasen2009,Tirel2017,Renoult2018} and has also been observed in the stretching of capillary bridges using extensional rheometers such as Capillary Breakup Extensional Rheometer (CaBER) \cite{Stelter2000,Stelter2002,Oliveira2005,Rodd2005,Arnolds2010,Campo-Deano2010,Bhat2010}. 
These studies have consistently evidenced the linear viscous-capillary thinning, the exponential polymeric thinning \cite{Kolte1999} and the existence of drops attached to a thin filament \cite{Shelley2001,Wagner2005,Ardekani2010,Aytouna2013,Dinic2015,Dinic2017,Deblais2018}, depending on the fluid properties.

A remarkable feature of the thinning of viscoelastic solution is the ability to form long and persistent filaments. 
Scanning electron microscopy observations \cite{Sattler2012} suggest the extensional flow is heterogeneous, with local variation of polymer concentration and localised pinching \cite{Roche2011}. 
Yet, the BOAS morphologie appears from the initial wavelength of capillary instability modified by the central fibre, which behaves as a solid core. 
The annular film becomes unstable resulting in drops along the filament.	
Inside a drop, the polymers are in relaxed state but in the fluid necks, they are in stretched state \cite{Sattler2012}.

From the numerical point of view, the so-called BOAS structure was reproduced using the slender body approximations \cite{Bousfield1986} for Oldroyd-B constitutive equations. The BOAS results as a subtle competition between viscosity, inertia, capillary and viscoelasticity \cite{Bhat2010}, that controls the growth of drops and the pinch-off.
The resulting 1D model was used to predict the drop dynamics of the BOAS including drop migration, oscillation, merging and draining \cite{Li2003}. 
More recent 2D axisymmetric simulations, again used the Oldroyd-B model \cite{Ardekani2010,Turkoz2018,Valette2019}, were able to show the contribution of the polymeric stresses within the filament during the thinning and the drop formation. Specifically, the axial polymeric stress components exhibit a large magnitude and a self-similar radial distribution in time.

Here, a new method based on the image analysis is developed, to calculate the extensional viscosity and to map strains. 
Homologous polymer mixtures or bi-disperse polymers at different concentration are tested covering a large range of dimensionless numbers. 
For the later stages of filament evolution, a diameter-space-time (DST) diagram is used to represent the drop dynamics.

\section{Experimental setup}

\begin{figure}
\centering
\includegraphics[width=1.0\columnwidth, keepaspectratio]{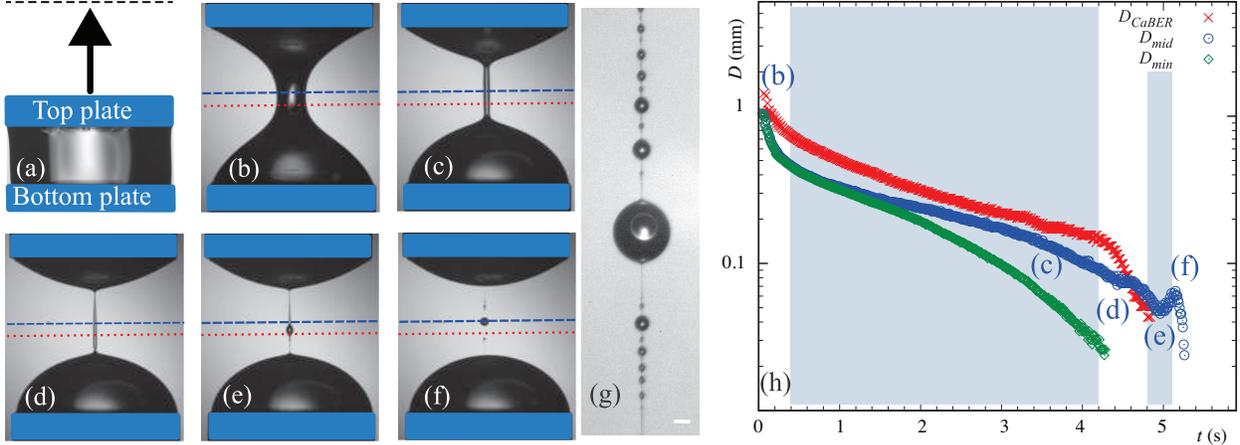}
\caption{(a-f) Photographs of the time-evolution of the  stretching of a capillary bridge until formation of drops for PEO2000. The fixed mid-height position of CaBER's micrometer is represented as a red dotted line and the mid-plane position of the filament is represented by a blue dashed line 
(g) Zoomed view of BOAS pattern, where the size of the big central drop is 300 $\mu$m. Scale bar: 100 $\mu$m. (h) $D$ versus $t$ for PEO2000, measured by the micrometer (red crosses) and by image processing at mid-plane of the filament, $D_{mid}$ (blue circles) and at the minimum diameter along the filament, $D_{min}$ (green diamonds). The shaded regions (b-f) represent the different thinning stages corresponding to photographs (b-f)}
\label{fig1}
\end{figure}

To measure the time-evolution of the filament diameter, CaBER's micrometer, as well as, high speed image processing have been used to capture the complete filament for measurement of the neck (pinch off) diameter and mid-plane diameter. The images are acquired at a rate from 100 to 1000 fps. 
The camera records stretching, filament thinning, pinch off, as well as, formation of drops on the filament. 
More details on the experiment and the protocols are described in supplementary material (see \footnote{See Supplemental Material at [URL will be inserted by publisher] for (i) the description of the materials and the fluid properties, (ii) space-time diagrams for all the fluids tested and (iii) two experimental videos for PEG20PEO1000 and PEG20PEO2000 are provided as PEG20PEO1000.gif (the complete filament thinning until breakup) and PEG20PEO2000.avi (the pinching and coalescence)}).
Poly-Ethylene Oxide (PEO), which is a high molecular weight polymer and Poly-Ethylene Glycol (PEG), which is a relatively low molecular weight, are used either separately or in combination with degassed deionised water. 
The relative high concentration of PEG, in PEG+PEO solution, makes solution majorly dependent upon PEG for shear viscosity and upon PEO for elasticity.
Broadly, our strategy is to control the shear viscosity of solution with PEG and the extensional viscosity with the concentration of PEO, noted $c_{PEO}$. 
The molecular weights of PEO and PEG are $8\times10^6$ and 20 000 g/mol, respectively.   
The solutions properties are in agreement with previous works \cite{Layec1983,Tirtaatmadja2006,Crumeyrolle2003}.
 
Fig. \ref{fig1}(a-g) depicts the time evolution of the filament stretching.
Initially, as shown in Fig. \ref{fig1}(a), a liquid bridge of 50 $\mu$L volume is formed in-between the two plates of CaBER with diameter, $D_0=6$ mm, separated by the initial height, $L_0 =2$ mm. 
The top plate is moved upwards linearly up to a final height, $L=6$ mm, over the time, $t = 50$ ms. 
Here, $t = 0$ is the instant when the top plate starts to move. 
The formation and thinning of filament until the formation of drops is represented in Fig. \ref{fig1}(b-f). 
After pinching of the filament at the both ends, the sample solution is pulled towards the centre of the filament, termed as recoil \cite{Li2003,Chang1999} as shown in Fig. \ref{fig1}(e). 
The recoiling further develops into drops-on-a-filament, as shown in Fig. \ref{fig1}(f), with a zoomed view of it in Fig. \ref{fig1}(g).
From the obtained images, with a self-developed code, the neck diameter or minimum diameter, $D_{min}$, and the diameter at filament's mid-plane, $D_{mid}$, are calculated for each image. 
In Fig. \ref{fig1}(h), the measured diameters are compared.
CaBER's micrometer measures the filament diameter, $D_{CaBER}$, at the fixed  mid-height between the plates, whereas, $D_{mid}$ is measured at the mid-plane of the filament, explaining the shift between the red and blue curves.
Clearly, the filament does not have a constant diameter over its length.
The $D_{mid}$ curve deviates from $D_{mid}$ from $t=2$ s.
Because of recoiling and coalescence, drops move axially along the thread, respective to mid-height, as well as, mid-plane, creating further crests and troughs in the later stages of the diameter versus time plot, as highlighted by the shaded regions (e) and (f) in Fig. \ref{fig1}(h). 

\section{Experimental results and discussion}

\begin{figure}
\centering
\includegraphics[width=1.0\columnwidth, keepaspectratio]{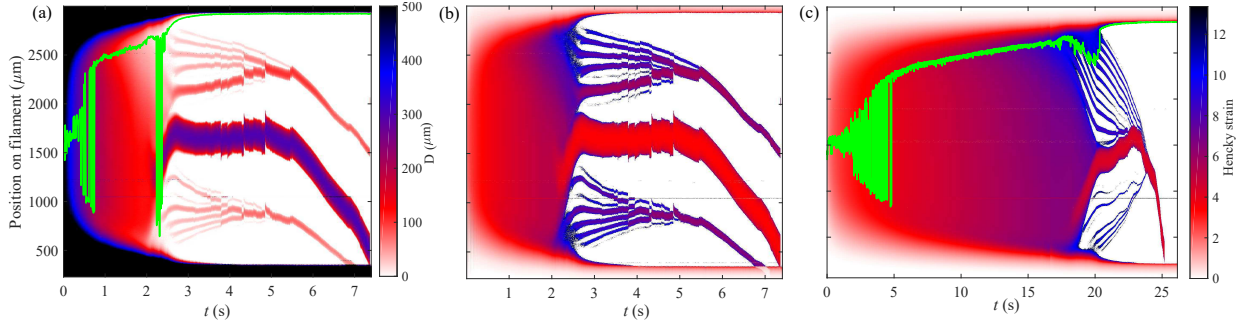}
\caption{(a) Diameter-space-time diagram, (b) Hencky strain-space-time diagram for PEO1000; and (c) Hencky strain-space-time diagram for PEG20PEO1000. Here, $t = 0$ is when the separation of the plates of CaBER starts. The green line represents the position of the minimum diameter of the filament.}
\label{fig2}
\end{figure}

The extension of polymer solution samples up to the breaking of filaments can also be represented through DST diagram and Hencky strain-space-time diagram, as depicted in Fig. \ref{fig2}.
The Hencky strain is defined as, $\epsilon(z,t) = 2 \ln \left[ D_0/D(z,t) \right] $ where $z$ is the vertical coordinate.
The diagrams are constructed from the images, where the diameter of the filament is converted into a colourcode, to obtain local diameter in space and time. 
The diagrams have been obtained for PEO1000 in Fig. \ref{fig2}(a) and (b), as well as, for PEG20PEO1000 in Fig. \ref{fig2}(c). 
The colourbars indicate the diameter, $D$, and  the Hencky strain, $\epsilon$, in space and time.  
It can be observed that for both solutions, a pattern emerges after the thinning of the filament, where the arrangement of drops on the filament proceeds with a relatively large drop forming around the mid-height with a diameter of 220-320 $\mu$m, accompanied by small drops on both sides.
At later time, for PEO solutions in Fig. \ref{fig2}(a), two relatively medium size satellite drops, with diameter varying from 50-150 $\mu$m, are observed on the either side of the big drop.
Interestingly, the diameter of the satellite drop on the top of the big drop is larger than the diameter of the satellite drop on the bottom.
The sizes of smaller drops, other than the big drop and two satellite drops, are less than 50 $\mu$m. 
In case of Hencky strain-space-time diagrams, it can be observed that $\epsilon$ is fairly uniform throughout the filament before pinching.
After $t = 2.5$ s for PEO1000 and $t=18$ s for PEG20PEO1000, drops on the filament appear and $\epsilon$ is higher in the smaller drops compared to the big central drop. 
In addition, the position of the minimum diameter, $D_{min}$, for each filament is calculated and represented by the green line. 
$D_{min}$ is initially located at the mid-height and then oscillates in between both ends of the filament. 
After pinching, $D_{min}$ is systematically located at the top of the filament, where the breakup occurs, presumably because of gravity effects.
From Fig. \ref{fig2}(c), it can be seen that $\epsilon$ corresponding to green line varies significantly over time after pinching.
Sudden vertical jumps in the diagrams are observed because of coalescence of small drops, for example in Fig. \ref{fig2}(a) at $t=4.3$ and 4.8 s.   
It suggests an elastic behaviour of the filament, as a result of the tension in the filament is relieved following every coalescence.
The motion of the filament occurs in the direction of coalescence. 
With increase in concentration of polymers, the thinning time, before the appearance of drops, and filament break-up time, which represents the time taken for filament to break, increase.  
Another feature of DST diagrams is the systematic fall of the big drop towards the bottom. 
The diagrams allow to measure the falling of the big drop. From the solution properties and the maximum diameter of the big central drop ($d_{PEG20PEO1000}=224$ $\mu$m), the calculated weight of the drop is 59.41 nN, assuming spherical shape. 
At equilibrium condition, the tension in the filament is equal to the weight of the drop. Then, the drop starts to fall. 
From Fig. 2(c), the equilibrium condition occurs at $t\simeq24$ s and at $z_0 = 1312$ $\mu$m, where $z_0$ is the vertical position of the drop on the filament from the bottom plate. The falling speed of the drop can then be measured; for PEG20PEO1000, $\mathcal{V}_{PEG20PEO1000} = z_{0} / \Delta t \simeq 1$ mm/s, where $\Delta t $ is time taken by the drop to fall into the solution pool on the bottom plate. 
In comparison, the velocity of the drop in free fall, $\sqrt{2gz_{0}} = 160$ mm/s, is higher than $\mathcal{V}_{PEG20PEO1000}$.
Hence, it can be concluded that there is non-negligible pulling force.
Also, it can be observed that the diameters of the drops in PEG+PEO solutions are smaller compared to the drops in PEO solutions.
It is interesting to notice that there are more satellite drops above the big drop than below, presumably due to gravity effects. 
The similar behaviour is observed for all PEO and PEG+PEO solutions and the additional diagrams are provided in [35].
The present space-time diagrams give the accurate measurements of the drops diameter and Hencky strain, whereas previous works \cite{Oliveira2005,Oliveira2006} are based on the grey-scale intensity to indicate relative thickness of the drops or the positions of the drops \cite{Clasen2009}.
Overall, the diagrams represent quantitative descriptions of the drop dynamics studied numerically by Li and Fontelos \cite{Li2003}, Ardekani {\it et al.} \cite{Ardekani2010} and Turkoz {\it et al.} \cite{Turkoz2018} using initial wave perturbation where filament thinning, drop migration, coalescence and draining were predicted.
Additionally, our experiments indicate the hierarchy and asymmetric distribution of satellite drops for all PEO and PEG+PEO solutions.

\begin{figure}
\centering
\includegraphics[width=1.0\columnwidth]{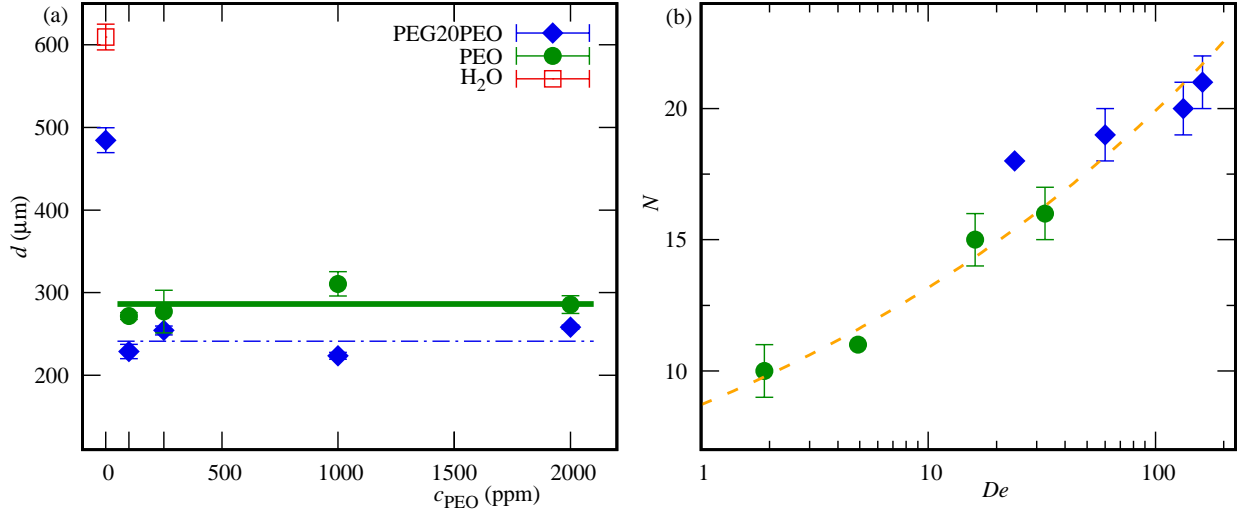}
\caption{Summary of the analysis of the diameter-space-time diagrams. (a) Diameter of the final big drop on the filament, $d$, versus PEO concentration. The thick and the dash-dotted lines represent the average final drop diameter. The error bars indicate the dispersion over 3 rehearsals. (b) Number of drops at the onset of the instability, $N$, as a function of $De$. The dashed line is power law fit with $N\propto De^{0.17}$.}
\label{fig3}
\end{figure}

Further insights of the drop dynamics from the diagrams can be obtained, as in Fig. \ref{fig3}(a), where the diameters of the final big drops on the filament for water, aqueous solutions of PEG, PEO and PEG+PEO are presented as a function of $c_{PEO}$. 
The error bars represent the dispersion over at least three repetitions of the experiment. 
For water and for PEG20, the diameters are in-between 600-620 and 480-500 $\mu$m, respectively. 
For the solutions containing PEO, the diameters of big drops decrease further in the range of 220-320 $\mu$m. 
It can also be observed that the average diameter of the big drops for the PEO solutions are comparatively higher than that of the PEG+PEO solutions.
However, the size of the big drop remains constant with increase in PEO concentration.
In Fig. \ref{fig3}(b), the maximum number of drops on filament after pinching is plotted against Deborah number, $De = \lambda / \sqrt{\rho D_0^3 / 8 \sigma}$ with $\lambda$ from CaBER data.
The dashed line is a power law fit predicting the maximum number of drops, $N$, varying as $De ^ {0.17}$, with the coefficient of determination 0.93.
The small value of exponent indicates limited increase in number of drops with $De$.
To the best of the author’s knowledge, there is no theoretical model available to compare this simple scaling law.
It is also interesting to note that the number of drops for PEG+PEO solutions are higher than PEO solutions for the same $c_{PEO}$.

\begin{figure}
\centering
\includegraphics[width=1.0\columnwidth, keepaspectratio]{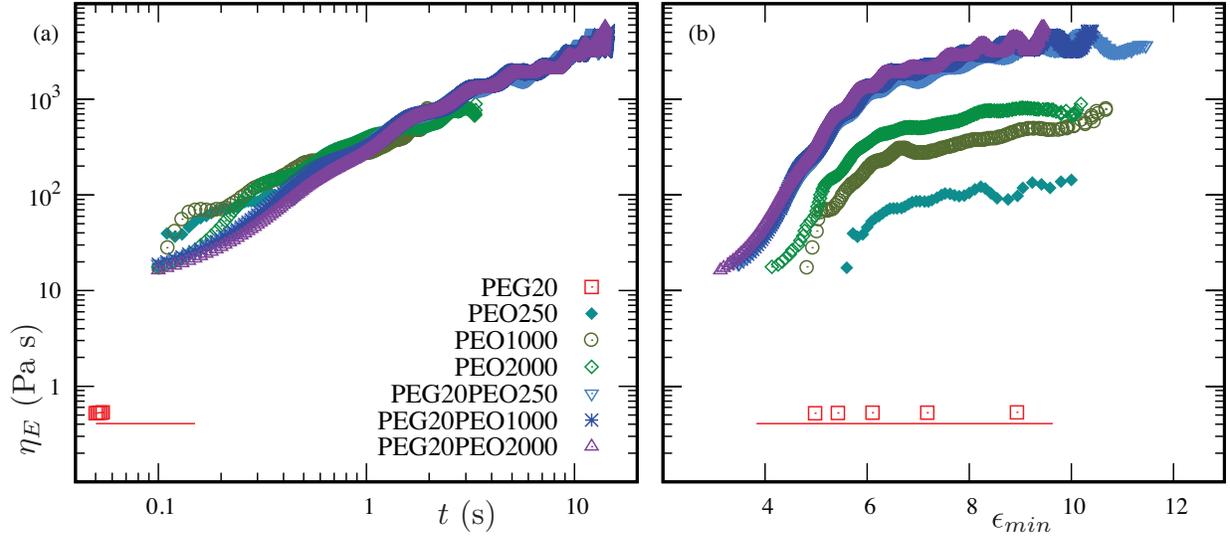}
\caption{(a) Extensional viscosity, $\eta_E$, for  PEG, PEO and PEG+PEO solutions as a function of time, $t$. (b) $\eta_E$ versus Hencky strain, $\epsilon_{min}$, at $D_{min}$. The red line indicates $\eta_E=3\eta_0$ for PEG20 from zero shear viscosity measurements.}
\label{fig4}
\end{figure}

\begin{figure}
\centering
\includegraphics[width=0.80\columnwidth, keepaspectratio]{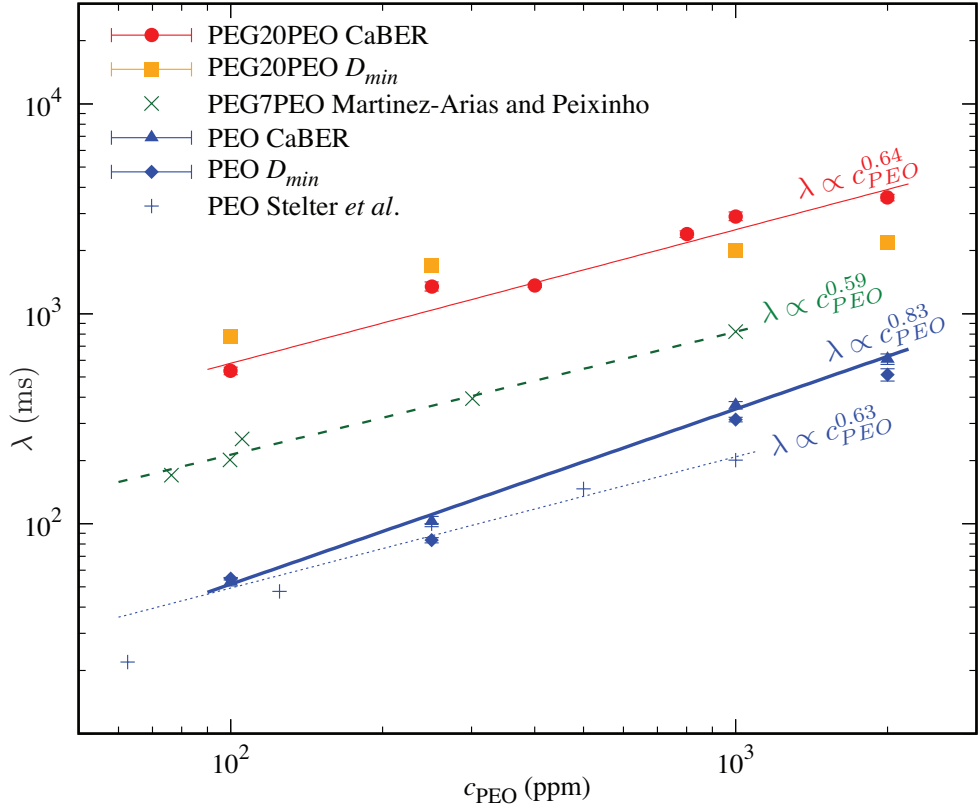}
\caption{The relaxation time, $\lambda$, as a function of the concentration of PEO, $c_{PEO}$, for PEO and PEG+PEO solutions. The graph represents comparison of $\lambda$ calculated by $D_{min}$ obtained from the image processing and from CaBER. Additionally, The lines indicate power law fits. The red thin line is a fit for PEG20PEO solutions and the thick blue line is a fit for PEO solutions. For both fits, $\lambda$ are measured with CaBER. The dark-green dashed line reproduces the fit of the results of Mart\'{i}nez-Arias and Peixinho \cite{Martinez-Arias2017} for PEG7PEO and the dotted line reproduces the fit of the results of Stelter {\it et al.} \cite{Stelter2002} for PEO solutions.}
\label{fig5}
\end{figure}

To gain more insight into the filament thinning, $D_{min}$, obtained from DST diagrams, can be used to calculate the apparent extensional viscosity, $\eta_E=-\sigma/(d{D_{min}}/{dt})$, where $\sigma$ is the surface tension.
The obtained results of $\eta_E$ for PEG, PEO and PEG+PEO solutions are plotted against time and Hencky strain, $\epsilon_{min}=2\,\ln \left[ D_0/D_{min}(t) \right]$, in Fig. \ref{fig4}(a) and (b), respectively.
The figures indicate an increase in $\eta_E$ with time, as well as, Hencky strain in agreement with previous studies {\cite{Yu2006,Gaillard2019}. 
Note that at pinching, $\eta_E$ tends to a constant, which seems to depend on the polymer concentration.
A closer look at the Hencky strain-space-time diagrams, for example Fig. \ref{fig2}(c), retrieves the Hencky strain corresponding to the green line that is $\epsilon_{min}$ of Fig. \ref{fig4}(b). 
From
the same figure,
it can be seen that all PEG+PEO solutions have comparatively higher $\eta_E$ than PEO solutions. 
Hence, with addition of PEG, $\eta_E$ of PEO solutions increases.
Gaillard {\it et al.} \cite{Gaillard2019} used the same molecular weight of PEG and PEO, as well as, minimum diameter of the filament to calculate $\eta_E$. 
Our results give the same $\eta_E$ for PEG+PEO solutions. 
In addition, Yu {\it et al.} \cite{Yu2006} used lower molecular weights of PEG and PEO. Hence, they found similar trends but lower values.
The change in $\eta_E$ with filament thinning can be characterised in two distinct regimes. 
The first regime, where the polymer chains are stretched, corresponds to development of the cylindrical shaped filament in the axial, as well as, radial direction. 
In this regime, for $\epsilon_{min}<6$, a small increase in $\epsilon_{min}$ results in large increase of $\eta_E$. 
Finally, $\eta_E$ transits to the second regime, where it does not increase much indicating strain hardening and the fully stretched state of polymers.
In this regime, the filament thinning is in the radial direction only.

Relaxation time, $\lambda$, is an important factor that quantifies viscoelastic fluids apart from Newtonian fluids.
$\lambda$ is calculated from ${D(t)}/{D_0}\propto\exp \left(-{t}/{3\lambda}\right)$ \cite{Kolte1999}. 
By fitting this equation in the elastocapillary regime, $\lambda$ is calculated and Fig. \ref{fig5} represents the change in relaxation time for different $c_{PEO}$ for PEG+PEO, as well as, PEO solutions on log-log scale. $\lambda$ has been calculated using CaBER, as well as, $D_{min}$. 
$\lambda$ calculated from $D_{min}$ for PEO solutions is in good agreement with $\lambda$ from CaBER.
It can be seen that $\lambda$ of PEG+PEO solutions increase with increasing $c_{PEO}$. 
Similar trend is observed for PEO solutions. 
Also, for the same $c_{PEO}$ solutions, $\lambda$ increases with addition of PEG. 
For example, relaxation time of PEG20PEO1000 solution is higher compared to the relaxation time of PEO1000 and PEG7PEO1000 solutions. 
The obtained data is compared with Stelter {\it et al.} \cite{Stelter2002} and Mart\'{i}nez-Arias and Peixinho \cite{Martinez-Arias2017}. 
Both used PEO with molecular weight of $8\times 10^6$ g/mol. 
The difference in $\lambda$ measured in the present results and  \cite{Stelter2002} can be because of differences in molecular weights of PEO and different protocols in the preparation of solutions. 
From Fig. \ref{fig5}, for the same $c_{PEO}$, $\lambda$ increases increases with the addition of PEG. For example, the relaxation time of PEG20PEO1000 solution is higher compared to the relaxation time of PEO1000 and PEG7PEO1000 solution. Hence, it can be concluded that higher the concentration of PEG in the aqueous solutions of PEO, higher will be the relaxation time of the solution. 

\section{Conclusions}
In conclusion, DST diagrams and the Hencky strain-space-time diagrams presented here allows to quantify the drop dynamics on the viscoelastic filament. 
By testing bi-disperse polymer solutions of various concentrations, it is found that there is a robust configuration of drops on the filament with a big central drop accompanied by smaller drops on both sides, with systematic movement of the filament in the direction of coalescence.
The diagrams also quantify filament thinning, drop migration, coalescence and draining, together with the position of the minimum diameter.
Interestingly, with addition of polymers in water, the size of the big central drop decreases significantly and then remains constant.
During the instability, the maximum number of drops increases with $De$ and can be predicted using a power law scaling.

\acknowledgements
The authors thank the Project BIOENGINE, which was co-financed by the European Union with the European Regional Development Fund and by the R\'{e}gion Normandie. JP would like to thank Rodrigue Mbakop for his help with preliminary rheological measurements. Our work also benefited from useful discussions with Prof. G\"{u}nter Brenn and Prof. Innocent Mutabazi.

\bibliographystyle{unsrt}
\bibliography{Filament}
\end{document}